\documentclass[sigconf, nonacm, 10pt]{acmart}
\setcopyright{none}
\renewcommand\footnotetextcopyrightpermission[1]{}
\AtBeginDocument{%
  }

\usepackage[acronym]{glossaries}
\usepackage{subcaption}
\usepackage{xcolor}

\makeglossaries
\glsdisablehyper

\newacronym{dash}{DASH}{Dynamic Adaptive Streaming over HTTP}
\newacronym{mdc}{MDC}{Multiple Description Coding}
\newacronym{vpcc}{V-PCC}{Video-based Point Cloud Compressiong}
\newacronym{gpcc}{G-PCC}{Geometry-based Point Cloud Compressiong}
\newacronym{abr}{ABR}{Adaptive Bitrate}
\newacronym{lru}{LRU}{Least Recently Used}
\newacronym{qoe}{QoE}{Quality of Experience}
\newacronym{cdf}{CDF}{Cumulative Distribution Function}
\newacronym{mpd}{MPD}{Media Presentation Description}
\newacronym{ar}{AR}{Augmented Reality}
\newacronym{vr}{VR}{Virtual Reality}

\begin{document}

\title{Scalable On-the-fly Transcoding for  Adaptive Streaming of Dynamic Point Clouds}
\author{Michael Rudolph}
\email{michael.rudolph@ikt.uni-hannover.de}
\orcid{0000-0001-5566-7783}
\affiliation{%
  \institution{Leibniz University Hannover}
  \city{Hannover}
  \country{Germany}
}

\author{Matthias De Fré}
\email{mattthias.defre@ugent.be}
\orcid{0009-0004-6044-5658}
\affiliation{%
  \institution{IDLab, Ghent University - imec}
  \city{Ghent}
  \country{Belgium}
}

\author{Finn Schnier}
\email{finn.schnier@stud.uni-hannover.de}
\orcid{0009-0002-2128-1545}
\affiliation{%
  \institution{Leibniz University Hannover}
  \city{Hannover}
  \country{Germany}
}

\author{Tim Wauters}
\email{tim.wauters@ugent.be}
\orcid{0000-0003-2618-3311} 
\affiliation{%
  \institution{IDLab, Ghent University - imec}
  \city{Ghent}
  \country{Belgium}
}

\author{Amr Rizk}
\email{amr.rizk@ikt.uni-hannover.de}
\orcid{0000-0002-9385-7729}
\affiliation{%
  \institution{Leibniz University Hannover}
  \city{Hannover}
  \country{Germany}
}


\begin{abstract}
On-the-fly transcoding of dynamic point cloud sequences reduces storage requirements and virtually increases the number of available representations for on demand streaming scenarios.
On-the-fly transcoding introduces, however, additional workload to media providers' infrastructure.
While V-PCC encoded content can be efficiently transcoded by re-encoding the underlying video bitstreams, which greatly benefits from hardware-accelerated video codec implementations, the scalability of such a system remains unclear.
In this work, we introduce and evaluate a dynamic point cloud streaming system that utilizes on-the-fly transcoding.
We explore the limits of scalability of this system in terms of request fulfillment times, specifically evaluating  the perceived user Quality of Experience.
We empirically show how caching and speculative transcoding allow to significantly reduce transcoding loads, allowing to scale to a higher number of simultaneous clients.
\end{abstract}




\maketitle

\section{Introduction}
Large scale delivery of immersive content for Augmented and Virtual Reality (AR/VR) applications introduces novel challenges to networked multimedia systems. 
Point clouds have emerged as a popular representation for immersive streaming, as they are easy to capture and to render~\cite{viola_vr2gather_2023} while allowing flexible, low-overhead processing such as view-dependent segmentation~\cite{subramanyam_user_2020} or sampling-based~\cite{de_fre_scalable_2024} quality adaptation.
However, point cloud representations of realistic scenes often consist of millions of points per frame, resulting in prohibitively large raw data rates.
Consequently, a wide range of compression methods~\cite{li_mpeg_2024, li_mpeg_2024-1} have emerged to support bandwidth-limited end-user devices.

While compression reduces the required bitrate, fluctuating network conditions necessitate flexible and scalable delivery mechanisms.
For on-demand streaming scenarios, \gls{abr} streaming enables scalable content distribution through client-side adaptivity while ensuring broad accessibility by relying on HTTP-based delivery.
Consequently, prior work has applied this concept to immersive multimedia, specifically to dynamic point cloud streaming~\cite{van_der_hooft_towards_2019, hosseini_dynamic_2018}.
In a nutshell, a point cloud video is divided into temporal segments which can be downloaded and decoded independently. 
Since only a fraction of the content is downloaded and buffered at the client, this ensures short start-up delays of streaming sessions while accounting for user interactions such as skipping or early stopping.
Scalability in terms of users is ensured through client-side quality adaptation.
Clients select from a set of representations for each segment, offering different trade-offs between rate and quality.
This adaptation is based on a current estimate of the network condition, the buffer-level as well as possible visibility estimates of the immersive content~\cite{van_der_hooft_towards_2019, subramanyam_user_2020}. 

Content providers offer a set of representations encoded at varying bitrates, which is termed bitrate ladder. 
Thus, clients can select one of the representations of the bitrate ladder for every segment, each offering a unique quality-rate trade-off.
Naturally, by offering more representations of the content in the bitrate ladder, more fine-grained adjustment and smoother quality transitions are possible.

It has been observed that only a small fraction of the streamed media is widely  popular, i.e., the distribution of the content views empirically follows a power law~\cite{cha2009analyzing}.
As the majority of content is not popular and might only be watched few times, this results \textit{in a waste of storage for holding many representations of the same segment}. 
On-the-fly transcoding~\cite{li_cost-efficient_2018} allows to address this problem: 
Instead of storing all representations on the content provider side, only a fraction (commonly the representation encoded at the highest rate) is stored.
Upon request, a representation at lower bitrate can be derived from the high rate representation through transcoding.  
This was shown for \gls{vpcc} encoded content using efficient transcoding techniques ~\cite{rudolph_transcoding_2025} that operate solely on the video bitstream of the encoded content.

In this work, we present an \textit{adaptive bitrate dynamic point cloud streaming system} based on the on-the-fly transcoding paradigm.
We experimentally explore the scalability of this system for increasing number of clients under limited transcoding resources.

\section{Related Work}
A key design decision in immersive content streaming systems is the selection of the transport protocol. In~\cite{mi2020demonstrating}, the TCP-based protocol \gls{dash} was employed, as the use case required reliable delivery of the complete data stream to all users. By contrast, in latency-sensitive scenarios, UDP-based protocols such as WebRTC and QUIC are commonly preferred, as they prioritize low end-to-end delay over strict reliability~\cite{gunkel2024vp9, haseeb_octavius_2025}.

Due to a high number of active users, immersive streaming systems may experience network congestion, which reduces the bandwidth available to each user~\cite{yang2022mobile}. 
Such bandwidth fluctuations can cause playback interruptions, forcing users to wait until subsequent frames are received~\cite{martinez2021identification}. 
These stall events substantially degrade the user’s overall \gls{qoe}. 
To mitigate this issue, adaptive streaming techniques are commonly employed, dynamically reducing content quality to avoid playback stalls~\cite{qiao2020beyond}. 
As demonstrated in \cite{taraghi2021intense}, quality degradation has a less detrimental impact on \gls{qoe} than the occurrence of stall events.
Many adaptive streaming techniques initially developed for traditional and 360-degree video streaming have been adapted to support dynamic point cloud streaming. 
In~\cite{hosseini_dynamic_2018}, the authors extended \gls{dash}~\cite{sodagar_mpeg-dash_2011} for point cloud content, using sampling strategies to derive multiple representations of dynamic point cloud scenes stored on a media server.
Equipped with this, the authors of \cite{van_der_hooft_towards_2019} proposed rate allocation methods for scenes with multiple dynamic point clouds to optimize visual quality in the user view port, which was later supported by objective and subjective quality studies in~\cite{van_der_hooft_objective_2020}. 
Similarly, other methods extend these concepts by tiling strategies for point clouds~\cite{subramanyam_user_2020}, enabling more fine-grained quality adaptation based on the user's view-port. Other approaches~\cite{wu2025p2vs, de_fre_scalable_2024} instead reduce encoding latency by creating multiple distinct quality representations through point clouds sampling, which can be combined at the receiver side. 
Furthermore, these systems can be extended with prediction algorithms that estimate the optimal quality representation in advance, thereby further reducing end-to-end latency~\cite{hu2023understanding, zhang2021buffer}.

Utilizing \gls{abr} streaming for delivering point cloud sequences necessitates efficient methods for preparing and compressing multiple quality representations.
Most preceding work~\cite{van_der_hooft_towards_2019, van_der_hooft_objective_2020, subramanyam_user_2020, heidarirad_vv-dash_2025} on on-demand point cloud streaming relies on the \gls{vpcc} standard~\cite{li_mpeg_2024-1}, as it offers unprecedented rate-distortion performance for dynamic point cloud sequences at the cost of high encoding latencies~\cite{bui2021comparative}. 
Although fast \gls{vpcc} codec implementations~\cite{freneau_uvgvpccenc_2025} have been proposed, encoding latency still remains a concern.
This restricts applications to on-demand use-cases, where content is pre-encoded in offline-manner where latency is less critical.
Consequently, on-the-fly transcoding, as explored for video streaming~\cite{li_cost-efficient_2018}, remains challenging. 
This stands in contrast to a survey on volumetric video streaming's state-of-the-art~\cite{viola_volumetric_2023}, which underlines the necessity of transcoding in streaming systems. 
Similarly, the authors of \cite{liu_point_2021} emphasize the need for low-latency, low-computation transcoding for volumetric video streaming systems. 
Instead of fully encoding and decoding \gls{vpcc} encoded content, solely transcoding the video sub-streams in already encoded content ~\cite{rudolph_transcoding_2025} offers reduced complexity and allows to bridge this gap. 

\section{Approach}
\begin{figure}
    \centering
    \includegraphics[width=\columnwidth]{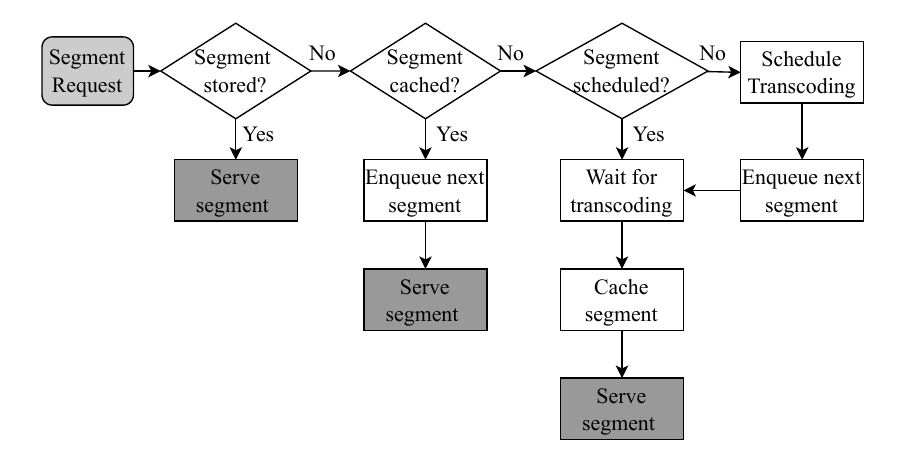}
    \vspace{-10pt}
    \caption{The transcoding decision process: Once the server receives a segment, it either serves from storage or forwards the request to the backend, which manages the transcoding job.}
    \label{fig:flowchart}
\end{figure}

\subsection{System Description}

Next, we present an on demand streaming system for HTTP adaptive streaming. 
The transcoding service consists of a media server frontend and a transcoder backend, with the decision flow for handling requests being depicted in Fig.~\ref{fig:flowchart}.
In addition to the
transcoder, we propose two extensions which aim at reducing the number of required transcoding jobs.
First, we introduce an \gls{lru} cache for transcoded segments, i.e. the segment with the oldest mark gets evicted. 
Further, we already enqueue the segment $n+1$ of a sequence upon the reception of request $n$.
This is rooted in the \gls{qoe}-tuning assumption that clients only switch qualities when necessary, and generally aim to keep streaming segments of the same quality. 

Fig.~\ref{fig:flowchart} summarizes the per-segment decision flow. 
Upon receiving a segment request, the server provides the requested segment from storage if available; otherwise, it forwards the request to the backend. 
After validating the request, the backend first checks the cache. 
If the segment is cached, a predictive transcoding job for the next segment is scheduled, and the requested segment is returned.
If not, the backend checks if a transcoding job for the same segment is already in progress.
If so, it waits for its completion; otherwise, it schedules a new transcoding job.
For both cases, speculative transcoding is triggered for the following segment.
Once the request segment finishes transcoding, it is added to the cache, and the backend responds to the request with the transcoded segment.

\subsection{Experimental Setup}
\begin{figure}
    \centering
    \includegraphics[width=\linewidth]{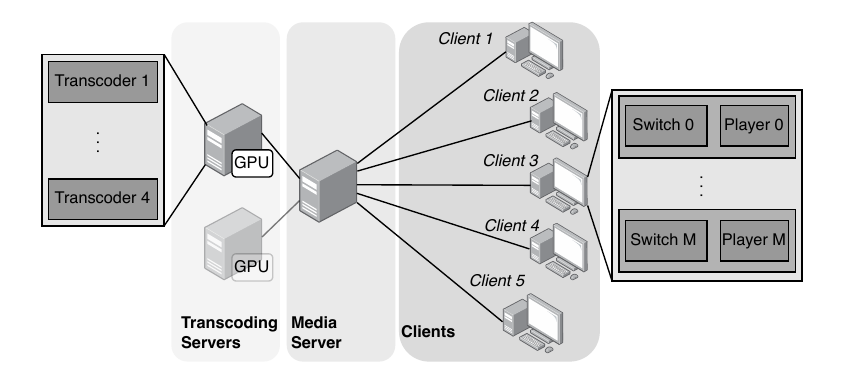}
    \vspace{-10pt}
    \caption{Experimental Setup: A central node serves as the media server. We use 1 or 2 nodes (indicated by lower opacity) for the transcoding backend and 5 client nodes to simulate a configurable number of players each. Each transcoding node in the backend is equipped with a GPU and supports 4 parallel transcoders. For player simulation, we route traffic through a software switch container, which emulates a network trace from~\cite{van_der_hooft_http2-based_2016} per player.}
    \label{fig:experimental_setup}
\end{figure}

We perform experiments on the  imec Virtual Wall~\cite{imec_virtualwall} testbed using a total of 8 bare-metal compute nodes.
The experiment topology is depicted in Fig.~\ref{fig:experimental_setup}.
A central media server, fronted by \texttt{NGINX}, handles incoming media requests in a \texttt{FastAPI} backend and distributes transcoding jobs via a \texttt{Redis}-backed job queue.
Containerized GPU-enabled worker nodes asynchronously consume jobs and independently transcode \gls{vpcc} bitstreams using the RABBIT transcoder~\cite{rudolph_transcoding_2025}.

We use client containers based on the iStream \gls{dash} Player~\cite{ansari_istream_2023}. 
The player is configured with a target buffer of 12~s, a safe buffer level of 8~s, and a panic buffer level of 2~s. 
The minimum buffer duration to resume after a stall is set to 2~s, and the startup buffer threshold is 3~s. 
A control script randomly selects one of the 4 encoded point cloud videos for each streaming session. 
Start times of streaming sessions within an experiment follow an exponential distribution with parameter $\lambda$ = 0.1 to avoid synchronized launches. 
Per-client bandwidth is shaped using a switch container, using \texttt{tc} to emulate a link according to bandwidth traces given in~\cite{van_der_hooft_http2-based_2016}, where each client is assigned an inidividual bandwidth trace.
We find that these traces offer a good fit to the bitrate ladder for the encoded test sequences, creating challenging conditions that force clients to use the full range of the offered representations. 

The experimental evaluation uses all four sequences from the 8iVFBv2 dataset~\cite{deon_8i_2017} which we extend to 80 seconds by looping the frames.
The extended sequence is encoded using the R5 \gls{vpcc} configuration defined in~\cite{isoiec_jtc_1sc_29wg_11_common_2020}.
The transcoder workers use \texttt{nvdec}/\texttt{nvenc} for re-encoding the geometry and attribute bitstreams in the pre-encoded bitstreams to the rate configurations R1-R4~\cite{isoiec_jtc_1sc_29wg_11_common_2020} using the \textit{p2} preset.

We run each streaming experiment for 600 seconds, using 1 or 2 GPU nodes for the backend. 
Since each GPU node is equipped with NVIDIA GeForce GTX1080 graphics card, we can run 8 simultaneous video encoders, resulting in 4 transcoding workers per node.
\gls{dash} clients are distributed on the physical client nodes, where each physical client node runs 4 or 8 \gls{dash} clients.
Each \gls{dash} client uses their own bandwidth emulation.

We compare the following experiment variants:
\begin{itemize}
    \item \textbf{Baseline (B)}: As a baseline for our experiments, we pre-encode all representations and store them on the media server. Consequently, no transcoding is required.
    \item \textbf{Transcoding (T)}: This variant performs transcoding upon request, without additional caching, pre-encoding or predictive encoding.
\end{itemize}

Further, we explore the following modifications to the transcoding system:
\begin{itemize}
    \item \textbf{Caching (+C)}: The media server has a small cache (128 MB) allowing to store a  portion of transcoded segments for direct response. 
    \item \textbf{Predictive Transcoding (+P)}: Once a segment is requested, we not only trigger a transcoding job for this segment, but also for the consecutive segment with the same coding configuration.
    \item \textbf{Fallback Pre-Encoding (+F)}: Instead of only pre-encoding and storing the highest rate representation, we additionally store the lowest rate representation on the media server. Since clients fallback to lowest rate representations when their buffer level falls below a threshold, this allows immediate server response without the need for a preceding transcoding job.
\end{itemize}

\section{Evaluation}

\begin{figure}
    \centering
    \includegraphics[width=\columnwidth]{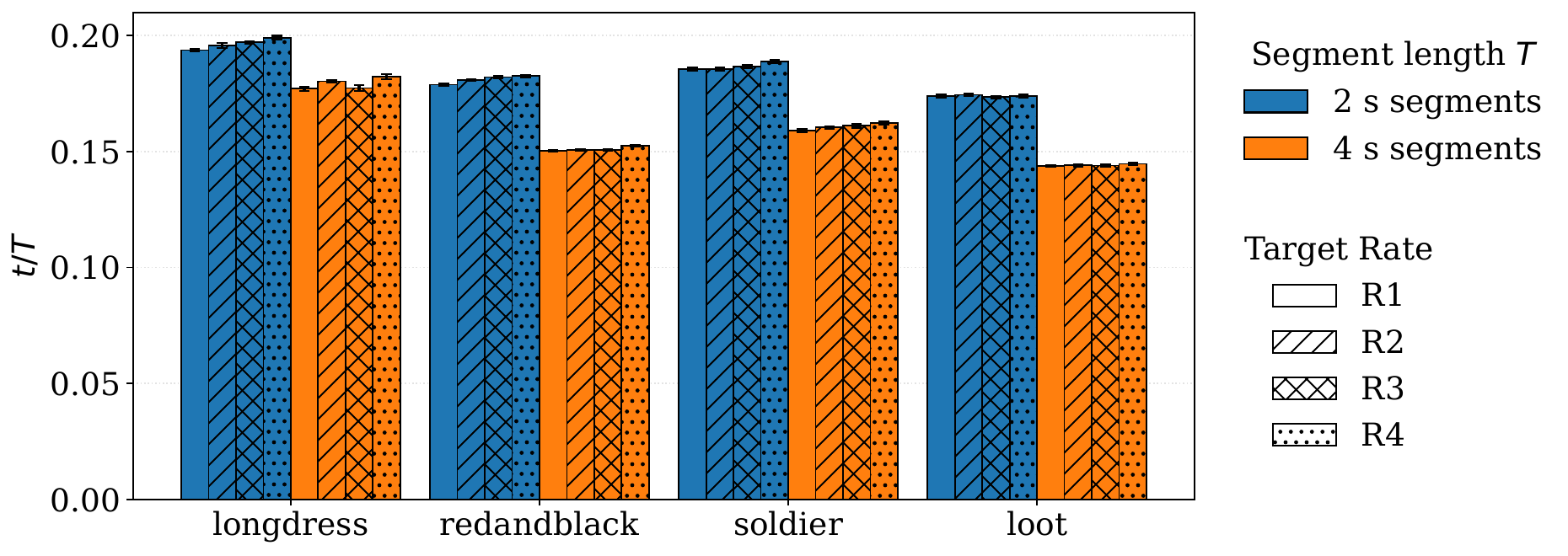}
    \vspace{-10pt}
    \caption{Transcoding time $t$ normalized by segment length $T$ for transcoding from rate configurations R5 to R1-R4. Averaged over 20 consecutive transcoding tasks of each looped sequence. Confidence intervals remain small.}
    \label{fig:transcoding_times}
\end{figure}

\subsection{Server-side Response Times}
Since on-the-fly transcoding requires processing of the request media before it can be served, it introduces additional response latency.
We visualize isolated time measurements of the transcoding latency on the testbed in Fig.~\ref{fig:transcoding_times}, which were averaged over 20 runs. 
To allow comparison between different segment lengths, we compute the normalized transcoding times  by dividing the measured transcoding times $t$ by the segment length $T$.
Although longer segments tend to result in lower normalized transcoding times $t/T$ (i.e. higher transcoding throughput), both segment lengths considered for our experiments can be transcoded sufficiently fast for on-the-fly transcoding.

\begin{figure}
\centering
\begin{subfigure}{0.23\textwidth}
    \includegraphics[width=\textwidth]{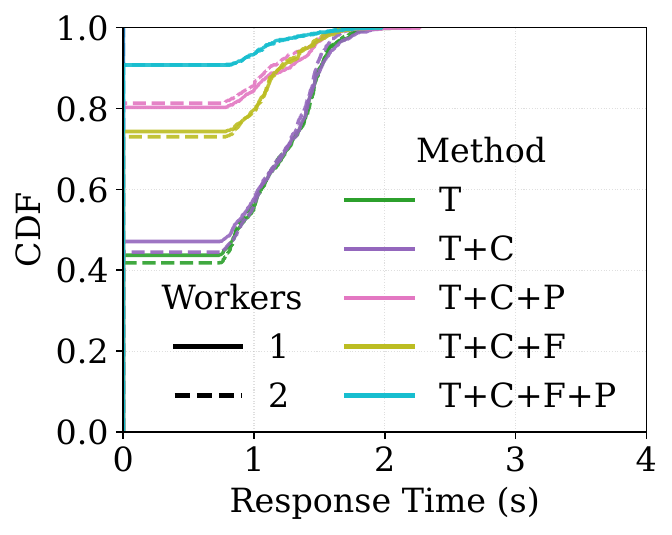}
    \vspace{-10pt}
    \caption{4 Clients, 2~s segments}
    \label{fig:cdf4_2s}
\end{subfigure}
\hfill
\begin{subfigure}{0.23\textwidth}
    \includegraphics[width=\textwidth]{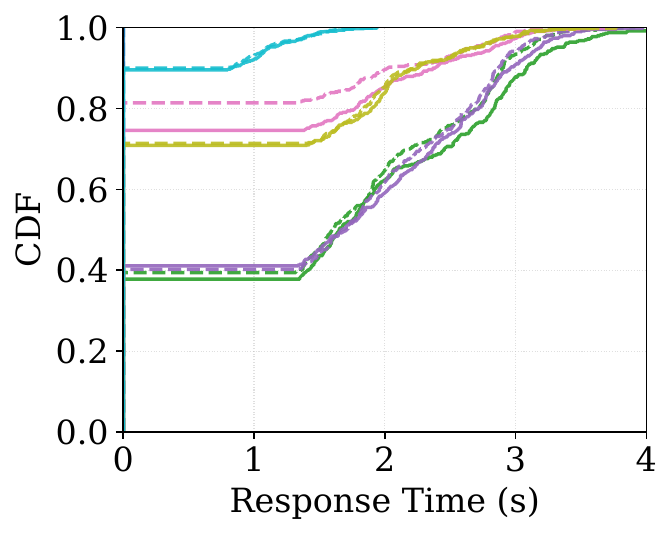}
    \vspace{-10pt}
    \caption{4 Clients, 4~s segments}
    \label{fig:cdf4_4s}
\end{subfigure}
\begin{subfigure}{0.23\textwidth}
    \includegraphics[width=\textwidth]{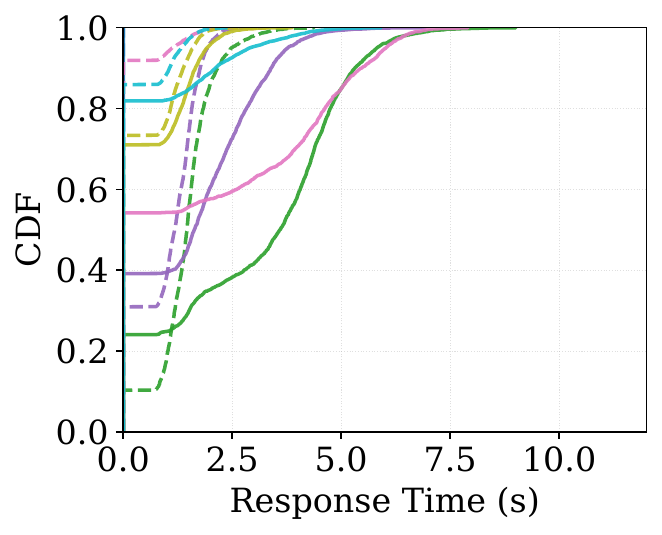}
    \vspace{-10pt}
    \caption{24 Clients, 2~s segments}
    \label{fig:cdf24_2s}
\end{subfigure}
\hfill
\begin{subfigure}{0.23\textwidth}
    \includegraphics[width=\textwidth]{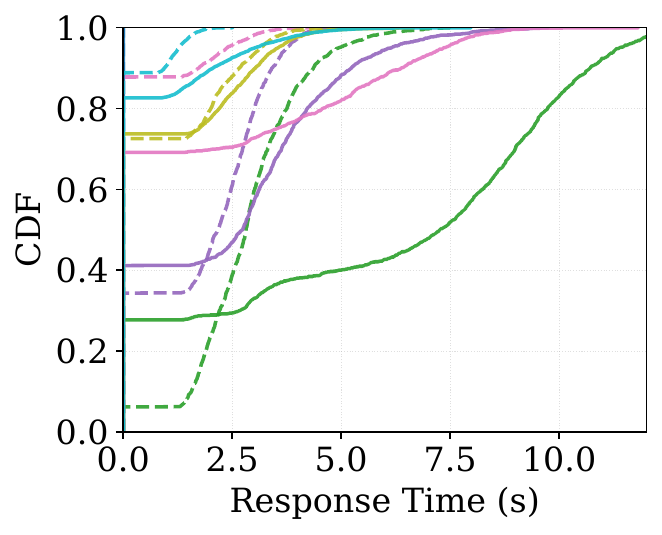}
    \vspace{-10pt}
    \caption{24 Clients, 4~s segments}
    \label{fig:cdf24_4s}
\end{subfigure}
\vspace{-5pt}
\caption{\glspl{cdf} of the response time (receiving a request until server-side response). The value of the CDF at zero latency (left axis) indicates the fraction of requests fulfilled instantaneously.}
\label{fig:repsonse_times}
\end{figure}

\begin{figure*}
    \begin{subfigure}{\columnwidth}
        \includegraphics[trim={0 0.3cm 0 0},clip, width=\textwidth]{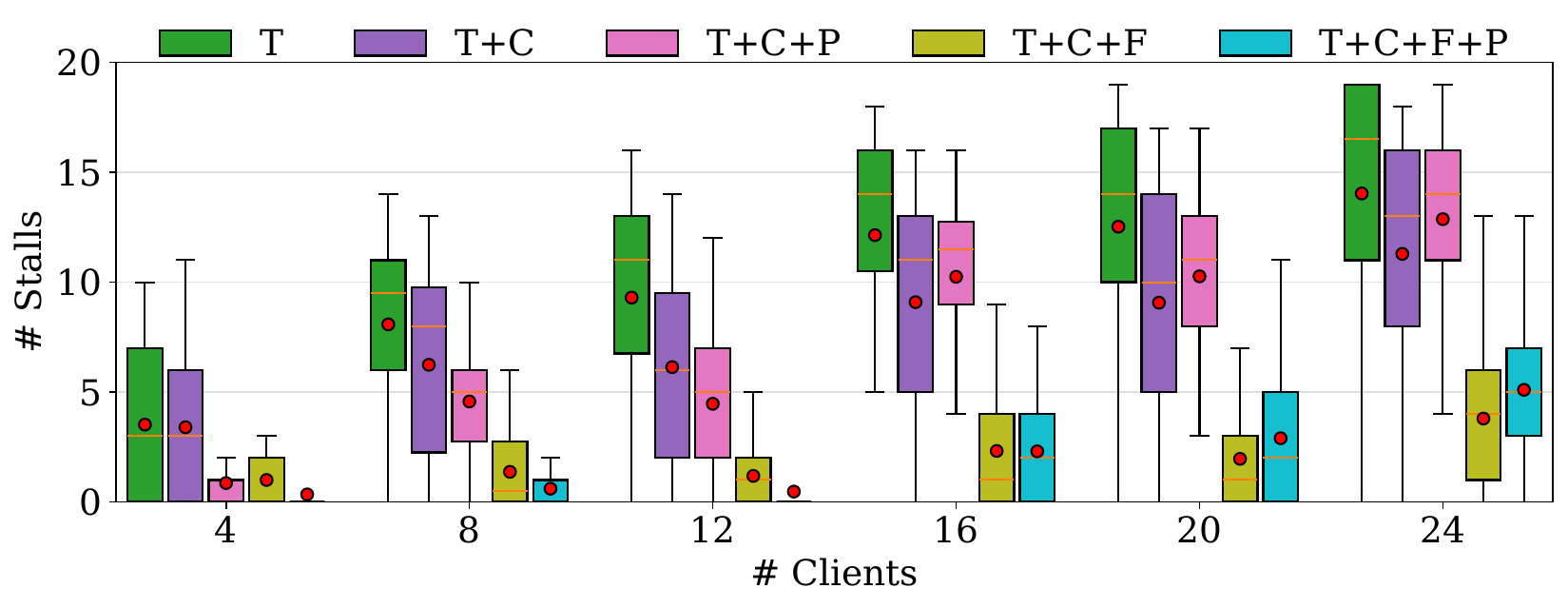}
        \vspace{-10pt}
        \caption{1 Worker, 2~s segments}
        \label{fig:stalls_w1_2s}
    \end{subfigure}
    \begin{subfigure}{\columnwidth}
        \includegraphics[trim={0 0.3cm 0 0},clip, width=\textwidth]{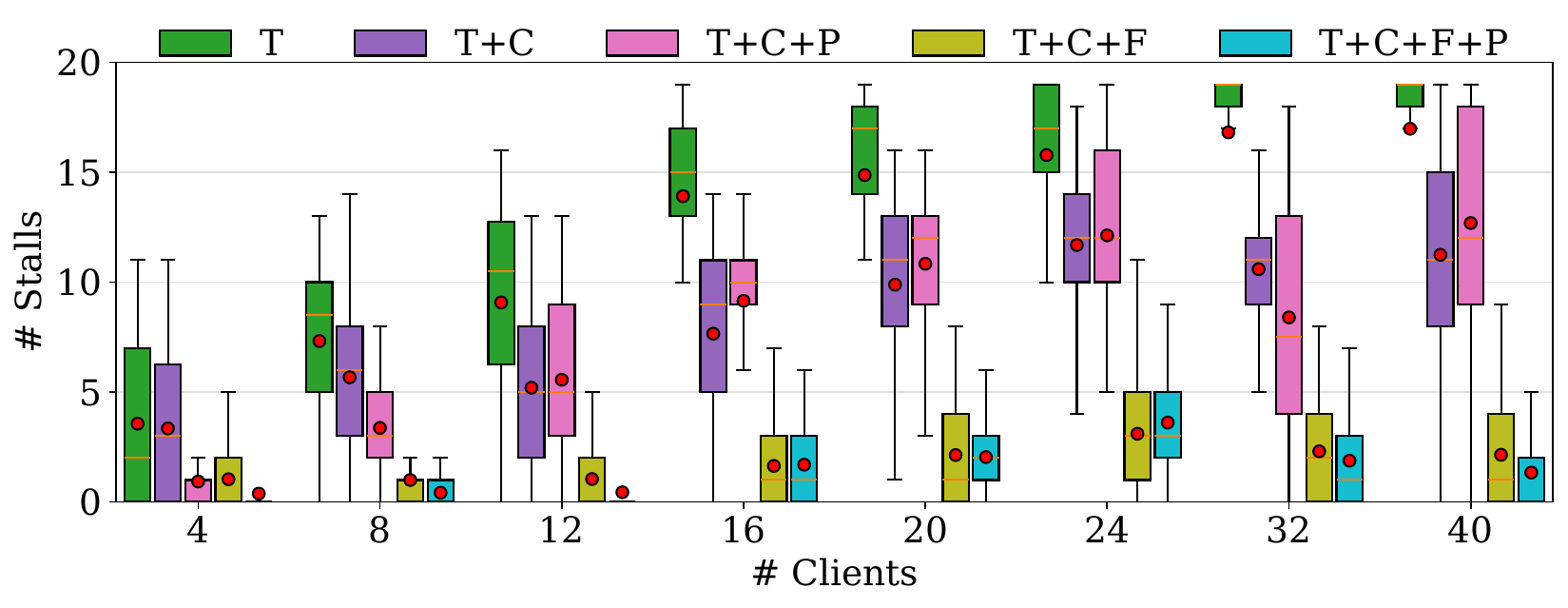}
        \vspace{-10pt}
        \caption{2 Workers, 2~s segments}
        \label{fig:stalls_w2_2s}
    \end{subfigure}
    \begin{subfigure}{\columnwidth}
        \includegraphics[trim={0 0.3cm 0 0},clip, width=\textwidth]{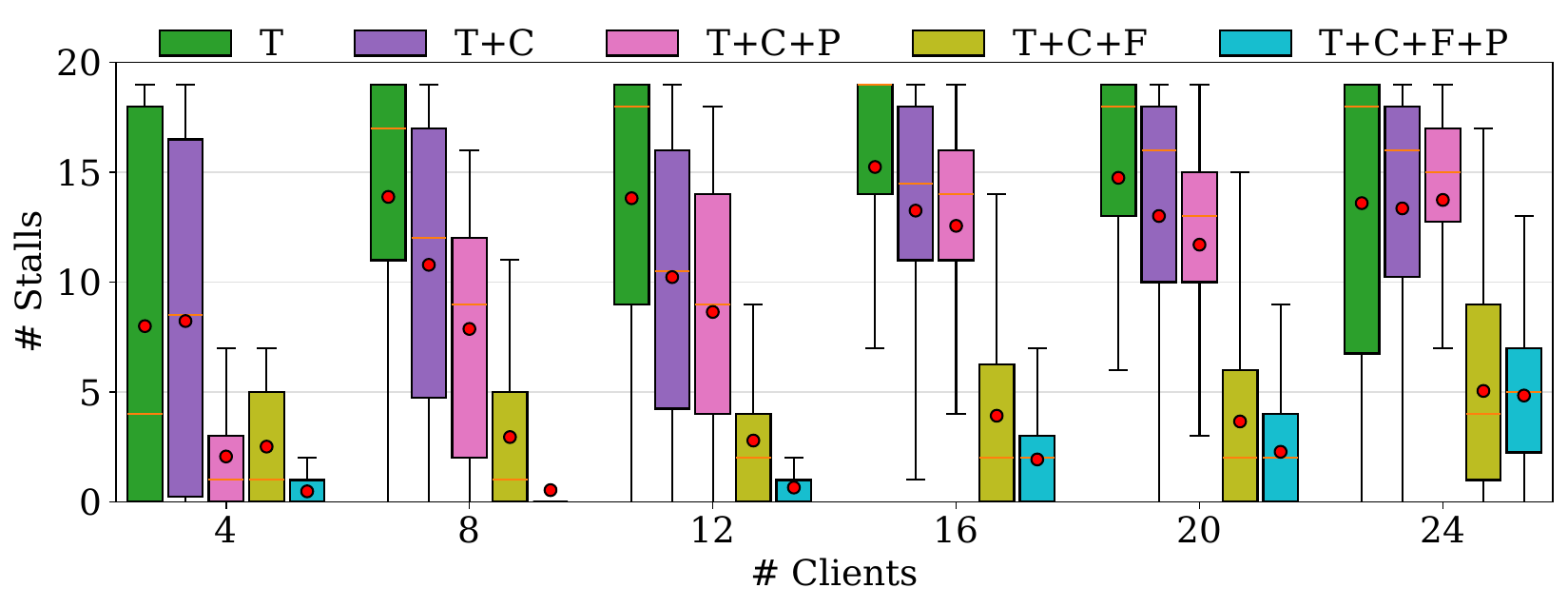}
        \vspace{-10pt}
        \caption{1 Worker, 4~s segments}
        \label{fig:stalls_w1_4s}
    \end{subfigure}
    \begin{subfigure}{\columnwidth}
        \includegraphics[trim={0 0.3cm 0 0},clip, width=\textwidth]{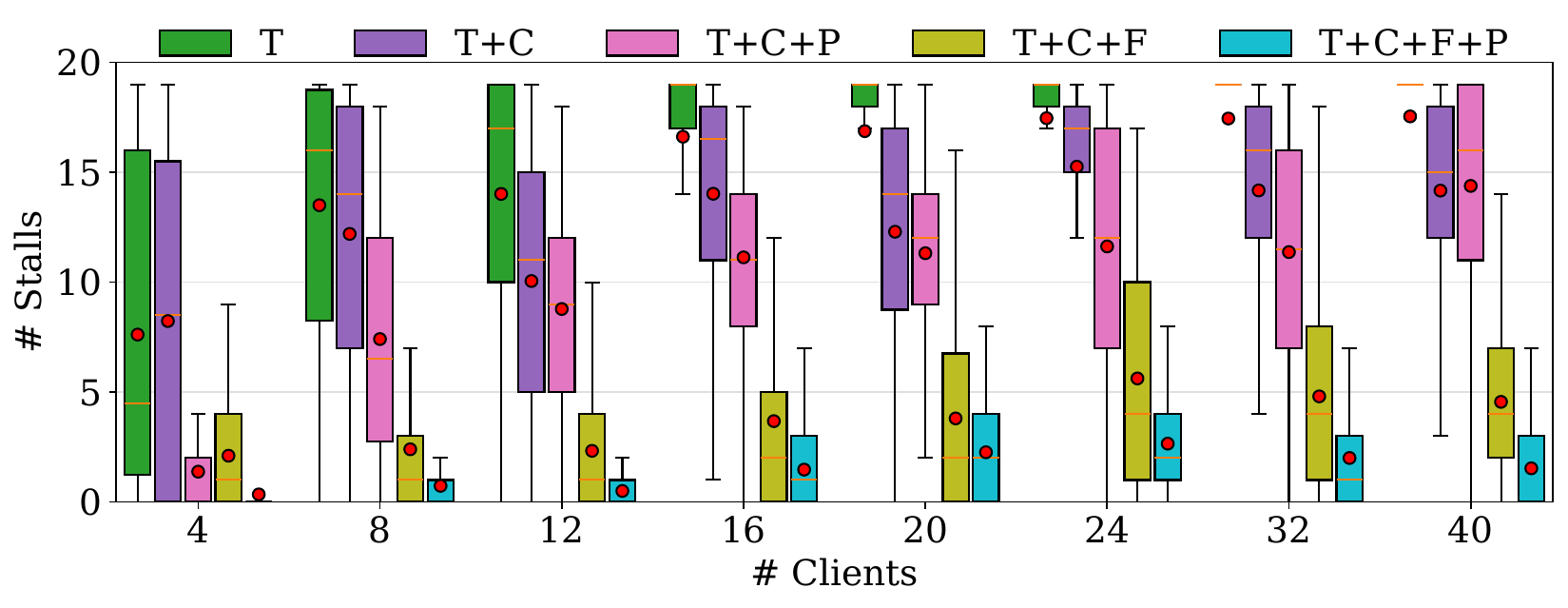}
        \vspace{-10pt}
        \caption{2 Workers, 4~s segments}
        \label{fig:stalls_w2_4s}
    \end{subfigure}
    \vspace{-6pt}
    \caption{Average number of stall events per streaming session: Pre-encoding the lowest quality representation as a fallback (F) proves most effectively in reducing stalls for increasing simultaneous client numbers, while Caching (C) proves more effective under higher loads. Predictive Encoding (P) reduces stall events for lower number of clients, but causes faster resource exhaustion and thus more stalls when the number of clients increases.}
    \label{fig:stalls}
\end{figure*}

From a systems perspective, the transcoding task latency alone only accounts for a part of the response latency, i.e. the time between a request arriving at the server until a response is sent back by the media server. 
Inspecting the \glspl{cdf} of the response time for the stand-alone Transcoder system (T; green) in a case with 4 clients in Fig.~\ref{fig:cdf4_2s} and Fig.~\ref{fig:cdf4_4s} reveals that the response time varies strongly for all requests received by the server.
A fraction of roughly 40\% of requests are for segments which are stored on the media server and can thus be resolved almost instantaneously.  
In contrast, the remaining requests trigger transcoding jobs, which result in variable response times through additional queuing delays for the transcoder pool. 
For this case, the difference between streaming experiments with 4 clients using 1 or 2 transcoding workers is small, indicating sufficient resources for both configurations. 
Introducing a cache (T+C) allows to significantly increase the proportion of requests which can be resolved directly, i.e. it reduces the number of transcoding jobs, as visible in Fig.~\ref{fig:repsonse_times}.
We observe a similar effect for predictive encoding (T+C+P), which triggers a speculative transcoding job for a consecutive segment at the same configuration, assuming the client won't switch the quality of the current stream, which further increases the proportion of direct responses.
In comparison, pre-encoding the lowest quality representation in the configuration T+C+F result in slightly less direct responses compared to T+C+P. 
Note, however, that this comes at the cost of additionally storing the lowest rate representation on the media server.
Finally, the system in configuration T+C+P+F only delays 10\% of the incoming request, while the remaining responses can be directly resolved from cache or storage.

For a larger number of simultaneous clients, and thus a higher rate of requests, we find increased server-side response latency as visible in Fig.~\ref{fig:cdf24_2s} and Fig.~\ref{fig:cdf24_4s}.
As transcoding jobs are processed by a limited number of workers due to resource limitations, bursts in requests result in additional queuing delays for jobs.
For 24 clients, the difference between using 1 or 2 transcoding workers is apparent.

\subsection{Client-side Quality}
The high variability of response latencies has a significant impact on the behavior of the streaming players.
High server-side response latency affects buffer filling of the clients, causing quality variations in buffer-filling based quality selection and eventually stall events when a player consumes segments from its buffer faster than it can be filled. 
Fig.~\ref{fig:stalls} shows the distribution of the number of stall events per streaming session for all experiment configurations.

\noindent\textbf{Only Transcoding:} We find that solely using the transcoder (T) results in a high number of stall events even for a low number of clients. 
This is independent of the number of workers used. 
High server-side response latency (cf. Fig.~\ref{fig:repsonse_times}), even for cases with sufficient resources, is already heavily impacting buffer management of clients. 

\noindent\textbf{Transcoding \& Caching (T+C):} An added cache (T+C) helps to slightly reduce the number of stalls for a low number of clients. 
The effectiveness of caching becomes more apparent with a higher number of clients, where the system with caching allows to effectively reduce the number of stall events compared to solely transcoding (T) for 40 clients (Fig.~\ref{fig:stalls_w2_2s} and Fig.~\ref{fig:stalls_w2_4s}).
At scale, caching allows to reduce the number of transcoding jobs and increase the proportion of requests that can be resolved directly.

\noindent\textbf{Transcoding, Caching \& Pred. Transcoding (T+C+P):} Adding the predictive encoding mechanism (T+C+P) reduces response times for a small number of clients when resources are sufficient.
However, once resources are exhausted and the job queue is filled, predictive encoding of consecutive segments becomes less effective, as it increases the transcoding work load and fills the job queue.

\noindent\textbf{Pre-endcoding the lowest quality (+F):} Finally, we find that pre-encoding the lowest quality representation (T+C+F and T+C+P+F) is most effective for\textit{ reducing the number of stalls} when increasing the number of clients.
Both variants strongly reduce the number of server-side on-the-fly transcoding jobs, allowing to scale significantly better with the number of clients. 

\begin{figure}
    \begin{subfigure}{\columnwidth}
        \includegraphics[width=\textwidth]{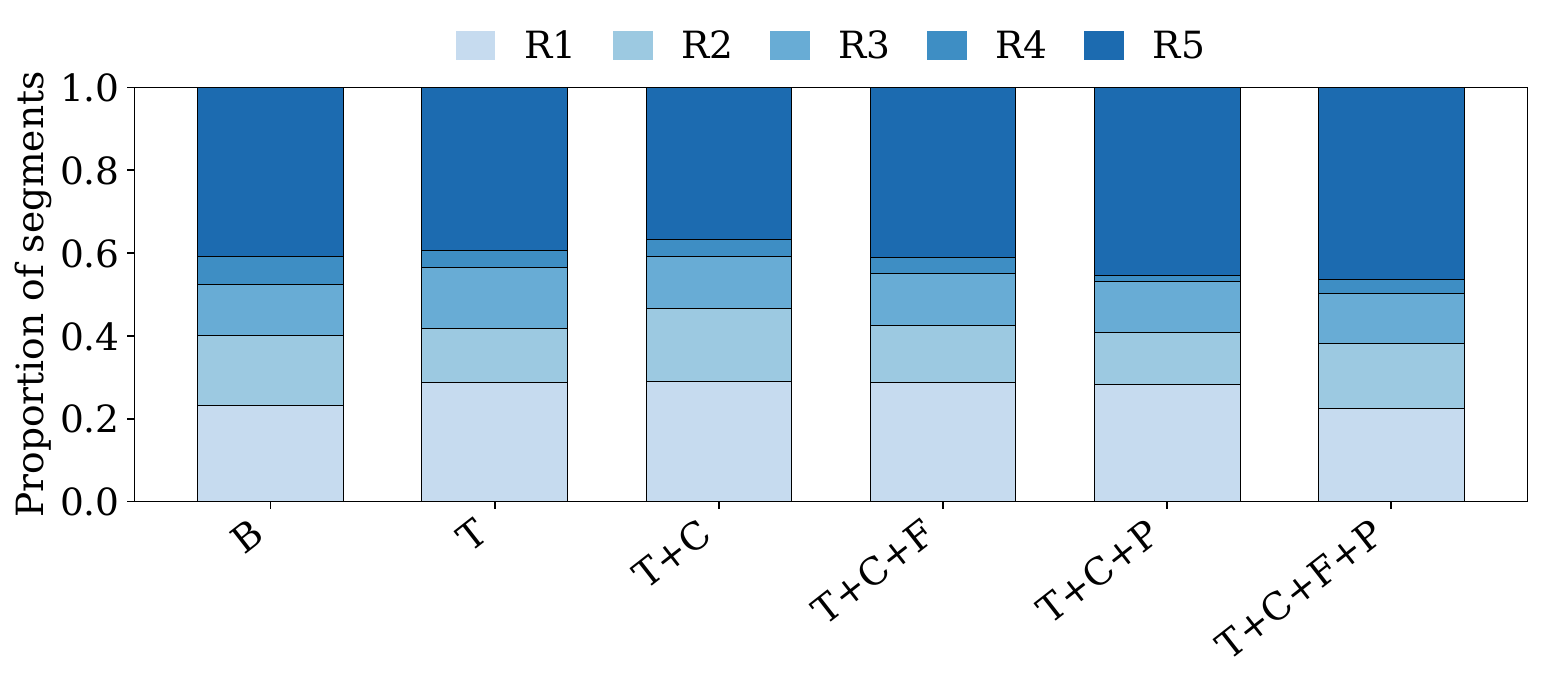}
        \caption{4 Clients}
        \label{fig:qual_4}
    \end{subfigure}
    \begin{subfigure}{\columnwidth}
        \includegraphics[width=\textwidth]{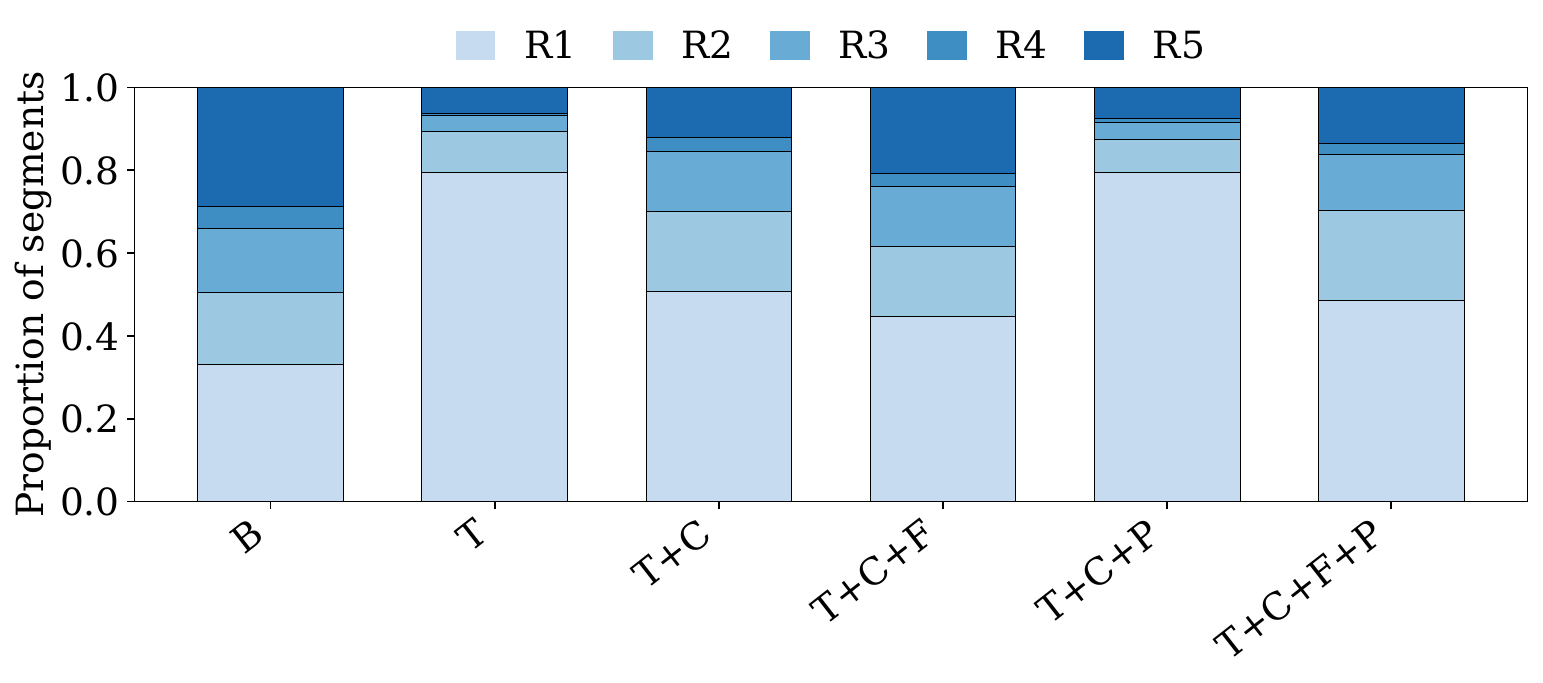}
        \caption{24 Clients}
        \label{fig:qual_24}
    \end{subfigure}
    \vspace{-10pt}
    \caption{Proportion of streamed segment qualities for the sequence \textit{redandblack} with 4~s segment length and 2 workers. The stacked bars indicate the fraction of received qualities over all \gls{dash} clients in the experiment with the mean representation indicated by the red line.}
    \label{fig:qualities}
\end{figure}

Since all variants of on-the-fly transcoding affect the selection of the clients' \gls{abr} algorithm, we illustrate the effect on the streamed rate representations, e.g. for the sequence \textit{redandblack} in Fig.~\ref{fig:qualities}. 
Quality selection for 4 clients is given in Fig.~\ref{fig:qual_4} for all on-the-fly transcoding configurations showing that it remains similar to the baseline (B) where all qualities are present at the server. 
In this setting transcoding shows a strong benefit where no need to keep all qualities present as in the baseline (B) and still the distribution of the streamed segment qualities is unchanged.
Recall that the clients run independent sample paths of the wireless network traces leading to a variation of the available bandwidth and hence the requested bitrates.  
Next, we observe quality degradation for 24 simultaneous clients in Fig.~\ref{fig:qual_24}.
We see that upon resource exhaustion, the players fail to increase their buffer to safe levels, preventing switches to higher rate representations.
The additional techniques, i.e. caching, predictive pre-encoding, and pre-encoding the lowest quality (+C, +P +F), respectively,  help improving the bitrate distribution to mimic the baseline, however, some performance loss is still apparent.  

\subsection{Discussion}
Under the challenging streaming conditions in our experiment, the added server-side response latency introduced by on-the-fly transcoding directly impacts the behavior of the \gls{dash} clients. 
Solely using a transcoder (T) without additional server-side measures to reduce the response time is insufficient for on-the-fly transcoding at scale, as it introduces severe stalling during client playback for an increasing number of clients. 

Assuming sufficient computing resources, predictive encoding combined with a cache (T+C+P) can substantially reduce the number of stalls by resolving a large fraction of requests directly at the server.
At large scale, however, speculative transcoding of segments that are never requested increases the load on the server at no benefit.

Finally, our results show that on-the-fly transcoding can provide the same streamed bitrate distribution as in the case when having all qualities present at the server. 
This would significantly reduce the requirement on storing dynamic point cloud scenes for adaptive bitrate streaming. 
As observed through the average number of stalling events and the distribution of the streamed bitrates, scaling on-the-fly transcoding with the number of clients significantly benefits from server-side helping measures such as caching, predictive pre-encoding and pre-encoding the lowest quality.

\section{Conclusion}
In this work, we presented and evaluated an on-the-fly trans\-coding system for adaptive bitrate streaming of dynamic point cloud sequences at scale. 
Our findings indicate that deploying a transcoder alone for the on-the-fly paradigm is insufficient specifically given small buffer filling targets and high network variability. 
To address this, we proposed system-level extensions to the transcoding system that substantially reduce the number of stalls compared to a transcoder-only approach.

When compared to a conventional streaming scenario where all representations are readily available on the server side, on-the-fly transcoding with server-side extensions is able to provide the same bitrate distribution for an increasing number of concurrent clients until resource exhaustion.
As the number of concurrent clients increases under limited transcoding resources we observe additional stalls and lower selected rate representations.

We identify several promising directions for future work:
Incorporating server-side quality decisions by predicting transcoding times and selectively responding with lower rate representation may prevent job queue buildup and excessive response times~\cite{hechler2024just}. 
Further, currently clients are unaware if a segment request can be served immediately or requires server-side processing. 
Exposing transcoding time estimates in the \gls{mpd} for transcoding-aware \gls{abr} algorithms could mitigate stalls at the cost of reduced visual quality.

\begin{acks}
Funded by the European Union (SPIRIT, 101070672). Views and opinions expressed are however those of the author(s) only and do not necessarily reflect those of the European Union. 
Neither the European Union nor the granting authority can be held responsible for them. 
The SPIRIT project has received funding from the Swiss State Secretariat for Education, Research and Innovation (SERI).
\end{acks}

\bibliographystyle{ACM-Reference-Format}
\bibliography{references}


\end{document}